\documentclass{PoS}
\usepackage{slashed}
\usepackage[utf8]{inputenc}
\usepackage{bbm}
\usepackage{setspace}
\usepackage{amsmath}
\def\Op{\mathcal{O}}

\def\l{\lambda}
\DeclareMathOperator{\Pf}{Pf}

\def\Tr{{\rm Tr}}

\DeclareMathOperator{\Di}{D}
\usepackage{stmaryrd}

\newcommand{\erw}[1]{\langle #1\rangle}
\newcommand{\arxiv}[1]{arXiv:\href{http://arxiv.org/abs/#1}{{\tt #1}}}
\title{Supersymmetric Yang-Mills theory: a step towards the continuum}

\ShortTitle{Simulations of supersymmetric Yang-Mills theory}

\author{\speaker{Georg Bergner}, Gernot M\"unster, Dirk Sandbrink, Umut D.\ \"Ozugurel\\
University of Münster, Institute for Theoretical Physics\\
Wilhelm-Klemm-Strasse 9, D-48149 Münster, Germany\\
E-mail: \email{g.bergner@uni-muenster.de}}
\author{Istvan Montvay\\
Deutsches Elektronen-Synchrotron DESY,\\
Notkestr. 85, D-22603 Hamburg, Germany}

\abstract{The spectrum of supersymmetric Yang-Mills theory presented so far shows an 
unexpected gap between the bosonic and fermionic masses. This finding was in 
contradiction with the basic requirements of supersymmetry. In this work we 
will present new results indicating that the mass gap is reduced at a smaller 
lattice spacing. Hence lattice artifacts are the most likely explanation for it. 
These new results have been obtained at a larger beta value and on a larger lattice.}

\FullConference{ The XXIX International Symposium on Lattice Field Theory - Lattice 2011\\
July 10-16, 2011\\
Squaw Valley, Lake Tahoe, California}

\begin{document}
\section{Introduction}
Supersymmetry is an important guiding principle for extensions of the standard model.
The symmetry relates fermionic and bosonic particles. 
Its algebra is a non-trivial extension of the Poincar\'e algebra by fermionic generators.
The anti-commutator of these generators is related to the generators of the translations.
A fundamental prediction of unbroken supersymmetry is a pairing of the energy states 
in the spectrum. 
Consequently  for each fermionic particle there exists a bosonic particle with the same mass.

A supersymmetric extension of Yang-Mills theory (super Yang-Mills theory) corresponds to 
the gauge sector of supersymmetric extensions of the standard model.
Fermionic fields (gluinos) $\lambda$ are added as fermionic partners of the gauge bosons.
The matching of the degrees of freedom in the bosonic and fermionic sector requires them to be
Majorana fermions in the adjoint representation of the gauge group.
The continuum Lagrangian has the following form 
\begin{equation}
 \mathcal{L}=\Tr\left[-\frac{1}{4} F_{\mu\nu}F^{\mu\nu}+\frac{i}{2}\bar{\lambda}\slashed{D}\lambda{-\frac{m_g}{2}\bar{\lambda}\lambda}\right]\, ,
\end{equation}
where $F_{\mu\nu}$ is the field strength tensor and the gauge group is $SU(N_c)$.
If the mass term ${ m_g}$ is zero supersymmetry is predicted to be unbroken, even in the quantized theory \cite{Witten:1982df}.
A second symmetry of this theory is the (chiral) $U(1)_R$ symmetry, $\lambda\rightarrow e^{-i\theta\gamma_5}\lambda$.
It is broken by the anomaly. The remaining $\mathbb{Z}_{2N_c}$ is spontaneously broken to $Z_2$.

The theory is assumed to be confining, therefore the low-energy degrees of freedom are formed from gauge invariant compound 
operators of the fields. The symmetries of the theory constrain the form of the low-energy effective action.
The first prediction for a low-energy effective action was derived from one supersymmetry multiplet that contains the meson
operators  $a-f_0$ ($\bar{\lambda}\lambda$), $a-\eta'$ ($\bar{\lambda}\gamma_5\lambda$), and the fermionic gluino-glue ($\sigma_{\mu\nu}F_{\mu\nu}\lambda$) \cite{Veneziano:1982ah}.
Later on, as an extension of the low energy effective theory, a multiplet of glueball operators ($0^{++}$ and $0^{-+}$) and the 
fermionic gluino-glue was added \cite{Farrar:1997fn}.
%
\section{Super Yang-Mills theory on the lattice}
Our approach is based on the work of Veneziano and Curci \cite{Curci:1986sm}.
The lattice action consists of the well-known plaquette action together with the
Wilson-Dirac operator $\Di_w$ in the fermionic sector \cite{Montvay:1995ea},
\begin{equation}
 \mathcal{S}_L=\beta \sum_P\left(1-\frac{1}{N_c}\Re U_P\right) +\frac{1}{2}\sum_{xy} \bar{\lambda}_x\left( \Di_w(m_g)\right)_{xy}\lambda_y\, .
\label{eq:lattac}
\end{equation}
The Wilson-Dirac operator is defined in the following way
\begin{equation}
 \Di_w=1-\kappa\sum_{\mu=1}^{4}
\left[(1-\gamma_\mu)_{\alpha,\beta}T_\mu+(1+\gamma_\mu)_{\alpha,
\beta}T^\dag_\mu\right]\, ,
\end{equation}
where $\kappa=\frac{1}{2(m_g+4)}$ and the $T_\mu$ lead to a gauge invariant shift of the 
fermion fields in $\mu$ direction
\begin{equation}
 T_\mu\lambda(x)=V_\mu \lambda(x+\hat{\mu})\, . 
\end{equation}
The links in the adjoint representation, $V_\mu$, are generated from the fundamental links using 
$(V_\mu)_{ab}=2\Tr[U_\mu^\dag T^a U_\mu T^b]$.

Supersymmetry and the $U(1)_R$ symmetry are broken in this discretization. 
Supersymmetry is known to be broken in any local interacting lattice theory \cite{Bergner:2009vg}. 
The $U(1)_R$ symmetry is, like chiral symmetry in QCD, broken by the Wilson mass. 
The breaking is not introduced in terms of a Ginsparg-Wilson relation, that ensures a modified symmetry on the lattice.
Hence a fine tuning of the parameters is needed to recover the symmetry in the continuum limit.
Fortunately it turns out that the number of terms, which must be considered in the fine tuning, is small -- in case of the lattice action \ref{eq:lattac} actually one. 

The Ward identities of supersymmetry and the $U(1)_R$ symmetry on the lattice are 
\begin{eqnarray}
 \erw{\nabla_\mu J_S^\mu(x)\, \Op(y)}&=&m_g\erw{D_S(x)\, \Op(y)}+\erw{X_{S}(x)\, \Op(y)}\\
 \erw{\nabla_\mu J_A^\mu(x)\, \Op(y)}&=&m_g\erw{D_A(x)\, \Op(y)}+\erw{X_{A}(x)\, \Op(y)}{+\propto\erw{F\tilde{F}\, \Op(y)}} \label{eq:xwi}.
\end{eqnarray}
The expectation values on the left hand side contain derivatives of the super-current $J_S$ and the current of the $U(1)_R$ symmetry $J_A$ with
an arbitrary operator composed from the fields of the theory.
A conserved symmetry requires the right hand side of these equations to be zero.
The terms $D_S$ and $D_A$ also appear in the continuum and represent the symmetry breaking induced by the non-zero gluino mass $m_g$.
The  $X_{S}$ and $X_{A}(x)$ are induced by the discretization and vanish in the classical continuum limit (tree level).
The last term in Eq.\ \ref{eq:xwi} stands for an additional contribution of the anomaly and is present also in the continuum.

In a quantized theory one has to include operators that can mix with $X_S$ and $X_A$ in the continuum limit.
The analysis of these operators for chiral symmetry and supersymmetry can be found in \cite{Bochicchio:1985xa} and 
\cite{Curci:1986sm}. In our case it leads to the following expressions
\begin{eqnarray}
 \erw{\nabla_\mu {Z_{A}}J^\mu_{A}(x)\, \Op(y)}&=&(m_g-{\bar{m}_g})\erw{D_A(x)\, \Op(y)}{+\propto\erw{F\tilde{F}\, \Op(y)}} + O(a)\\
 \erw{\nabla_\mu({Z_S} J_S^\mu(x)+{\tilde{Z_S} \tilde{J_S}^\mu(x)})\, \Op(y)}&=&(m_g-{\bar{m}_g})\erw{D_S(x)\, \Op(y)}+O(a) \, .
\end{eqnarray}
This means that up to effects that vanish in the continuum limit, the symmetry breaking of the discretization can be absorbed into a redefinition of the 
currents and a shift of the bare mass parameter $m_g$. A tuning of $m_g$ is hence enough to restore the chiral symmetry and supersymmetry in this theory.
Up to $O(a)$ effects the tuning for both symmetries is the same.

The above considerations show that the chiral and the supersymmetric Ward identities can in principle both be used to tune the bare mass parameter.
However, the signal of the supersymmetric Ward-identities does not allow for a precise tuning.
The chiral Ward-identities, on the other hand, contain an additional term from the anomaly.
To avoid these limitations, we consider the connected part (OZI approximation) of the $a-\eta'$, the adjoint pion ($a-\pi$).
The mass of this particle vanishes in the chiral limit. However, the $a-\pi$ can be defined only in a partially quenched setup.
Therefore, the consistency with the supersymmetric Ward identities must be verified in a separate measurement.
\section{Some details about the simulations}
\label{sec:tech}
\begin{figure}
\begin{center}
 \includegraphics[width=7.5cm]{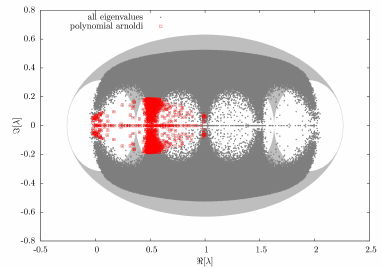}
 \includegraphics[width=7.5cm]{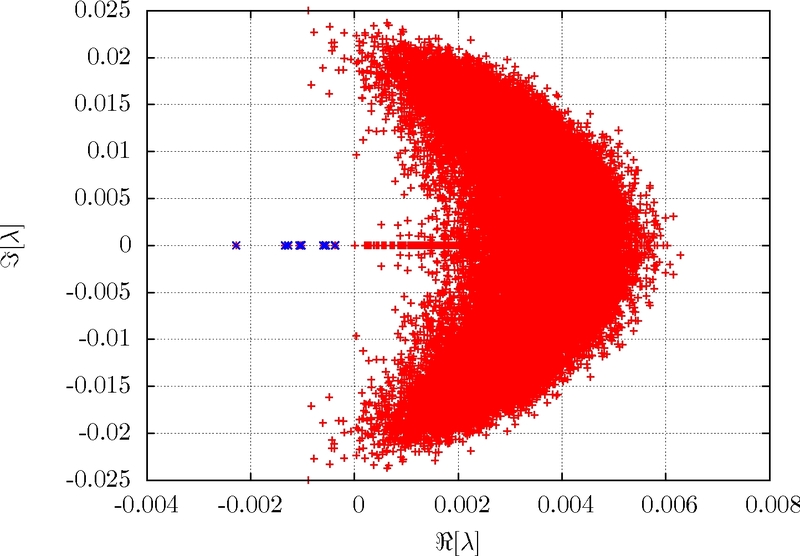}
\end{center}
 \caption{The left figure shows how our method focuses the calculation on the real eigenvalues. The red boxed points are the part of the complete spectrum of $\Di_w$ computed in the iteration. 
The 20 lowest eigenvalues in the right figure are obtained from 2007 configurations at $\kappa=0.1495$, $\beta=1.75$ on a $32^3\times 64$ lattice. Only 16 negative Pfaffians have been obtained. }
\label{fig:eigenv}
\end{figure}

The simulations of this theory include several specific challenges. The integration of the Majorana fermions leads to the Pfaffian $\Pf(C\Di_w)$, where $C$ is the charge conjugation matrix, instead of the determinant.
The modulus of the Pfaffian is the square root of the determinant. It can be included in the update algorithm using a polynomial approximation (PHMC algorithm \cite{Montvay:2005tj}).
The sign of the Pfaffian, however, can only be included in a reweighting of the measurements.
A detailed analysis shows that the Pfaffian has the same sign as the product of the doubly degenerate real eigenvalues \cite{Koutsoumbas:1997de}. If the number of negative real degenerate eigenvalues is odd, 
a negative Pfaffian sign must be included in the reweighting.
Our efficient way to calculate the real negative eigenvalues using the Arnoldi algorithm together with polynomial focusing and acceleration, illustrated in Fig.\ \ref{fig:eigenv}, can be found in \cite{Bergner:2011zp}.

\begin{figure}[h!]
 \begin{center}
 \includegraphics[width=9cm]{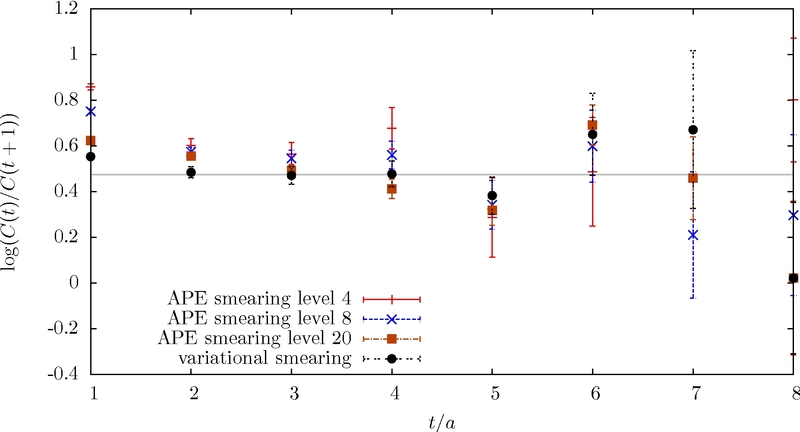}
\end{center}
\caption{The effective mass of the glueball obtained with several different methods. The variational smearing methods allow the determination of a plateau of the effective mass.
(This result has been obtained on a $32^3\times 64$ lattice, $\beta=1.75$, $\kappa=0.1492$, 2848 configurations.) }
\label{fig:glueb}
\end{figure}

The considered observables of the theory are the correlation functions of glueballs, mesonic operators, and the fermionic gluino-glue. 
A measurement of the mass of the $0^{++}$ glueball is possible using variational smearing methods (see Fig.\ \ref{fig:glueb}).

All of the meson operators in this model contain disconnected contributions, that become dominant in the chiral limit. 
There are several techniques to obtain these contributions like the  stochastic estimator technique (SET), 
or the improved volume source technique (IVST) \cite{Farchioni:2004ej}.
In our latest studies we have applied the SET  (see \cite{Bali:2009hu} for a review of this method), since it can be combined with the 
exact contribution of the lowest eigenmodes. In this method the inverse is approximated using random vectors ($\in \mathbb{Z}_{4}$).
We have improved the SET using the truncated solver method and the exact contribution of the lowest eigenmodes.

\begin{figure}[h!]
 \begin{center}
  \includegraphics[width=10cm]{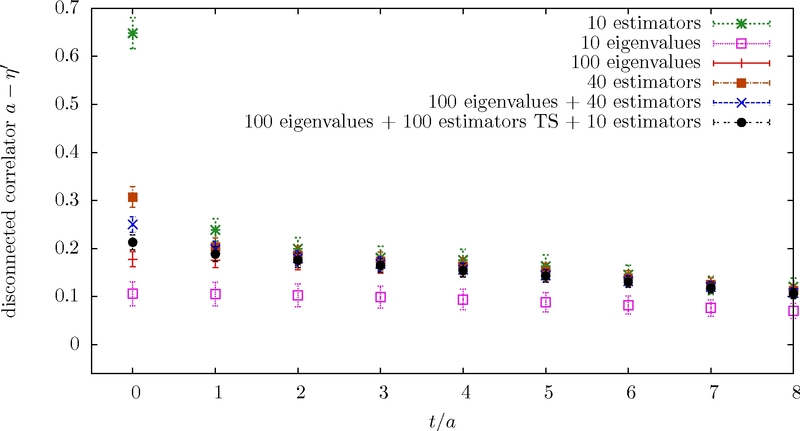}
\end{center}
\caption{This picture shows different strategies to obtain the disconnected contributions of the $a-\eta'$.
Different number of stochastic estimators in the SET are compared with a truncated eigenmode expansion using 10 and 100 eigenvalues. 
The last two strategies shown in this picture are a combination of the SET and  the truncated eigenmode expansion. In the last step a
truncated solver method (using 100 low precision and 10 precise estimators) has been applied. 
The projection of the noise to the space orthogonal to the lowest eigenmodes leads to a faster conjugate gradient inversion.
Therefore, the time needed for the last three strategies is comparable at small gluino masses.
(The disconnected correlator in this figure has been obtained on a $32^3\times 64$ lattice, $\beta=1.75$, $\kappa=0.1494$, on 300 independent configurations.)}
\label{fig:set2}
\end{figure}

An approximation of the inverse is obtained from the lowest eigenmodes of the Hermitian matrix $\gamma_5 \Di_w$.
When the noise is projected orthogonal to the eigenspace, this approximation can be combined with the SET.
In our case we have used the eigenvalues from the even-odd preconditioned Hermitian matrix. They have been obtained with the
Arnoldi algorithm using Chebyshev polynomials to accelerate the algorithm. The spectral decomposition using the eigenvalues of the 
even-odd preconditioned matrix combined with the SET and the truncated solver method leads to a considerable improvement for the determination 
of the $a-\eta'$ at small gluino masses. Note that the same eigenvalues are used to calculate reweighting factors to improve the polynomial approximation
in the PHMC algorithm. An illustration of these methods can be found in Fig.\ \ref{fig:set2}.

\begin{figure}
 \begin{center}
  \includegraphics[width=8.5cm]{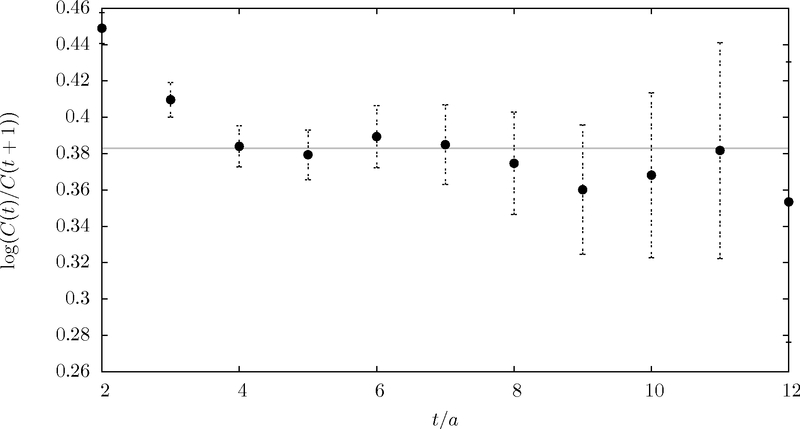}
\end{center}
\caption{The effective mass of the gluino-glue.
The correlator of the gluino-glue has been obtained with APE and Jacobi smearing. The gray line indicates the value obtained in a fit.
(This result has been obtained on a $32^3\times 64$ lattice, $\beta=1.75$, $\kappa=0.1492$, 2848 configurations.)}
\label{fig:gg}
\end{figure}

The fermionic particle considered in the mass spectrum is realized by the gluino-glue operator $\sigma^{\mu\nu}\Tr [F_{\mu\nu}\l]$, where $F_{\mu\nu}$ is represented by clover plaquettes.
A combination of APE smearing on gauge fields and Jacobi smearing on $\lambda$ has been used in this case. This combined smearing leads to a good signal compared to glueballs (see Fig.\ \ref{fig:gg}).

Our simulations are done with the PHMC algorithm with determinant breakup. 
A tree level Symanzik improvement of the gauge action and stout smearing for the fermion links has been added to the standard setup presented in the previous section.
It is expected that these improvements reduce the lattice artifacts. 

\section{Results}
\label{sec:results}
In our first simulations we have used the lattice sizes $16^3\times 32$, $24^3\times 48$, and $32^3\times 64$.
We have chosen $\beta=1.6$ and measured the Sommer scale ($r_0$). If one sets  $r_0\equiv 0.5\text{fm}$, like in QCD,
the lattice spacing is below $0.088\text{fm}$ and the lattice size is  around $1.5-2.3 \text{fm}$.
The smallest adjoint pion mass has been at about $440\text{MeV}$. 
The tuning of the bare gluino mass towards the chiral limit was found to be consistent with the supersymmetric Ward identities.
Details can be found in \cite{Demmouche:2010sf}.

The mass spectrum obtained in these simulations revealed an unexpected mass gap between the bosonic and fermionic particles. Such a mass gap is in contradiction 
to supersymmetry. Therefore a more careful analysis is essential.

As a next step of our scrutiny of the mass spectrum of supersymmetric Yang-Mills theory we have investigated the influence of the finite lattice spacing.
To decrease the lattice spacing we have increased $\beta$ from 1.6 to 1.75.
The simulations have been performed on a $32^3\times 64$ lattice.
The lattice spacing in these simulations is around $0.057\text{fm}$ with a lattice size of approximately $1.8 \text{fm}$.

\begin{figure}
\begin{center}
 \includegraphics[width=7.5cm]{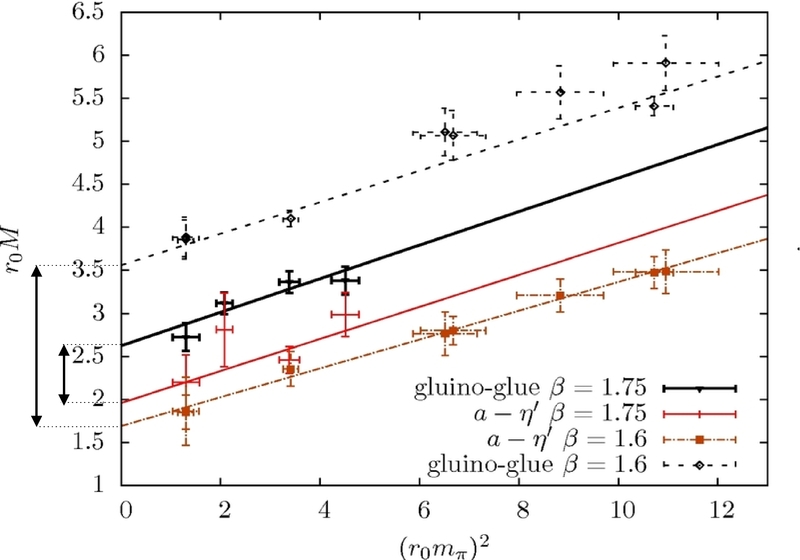} 
 \includegraphics[width=7.5cm]{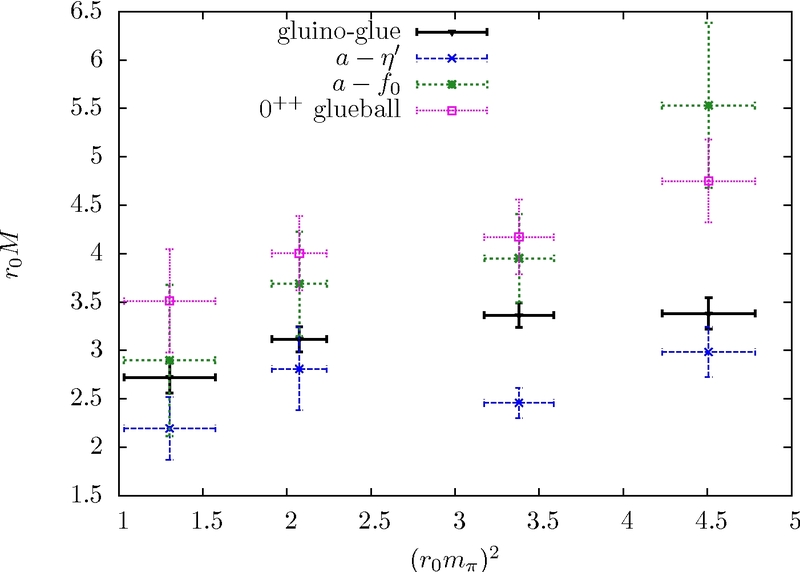} 
\end{center}
\caption{The mass spectrum at a smaller lattice spacing.
In the left figure the masses of the $a-\eta'$ and the gluino-glue are extrapolated to the continuum limit.
The values obtained at a smaller value of $\beta$ are included for comparison. The arrows indicate the mass gap in the chiral limit.
The right figure shows the masses of all particles obtained at $\beta=1.75$.
As expected the data suggest a mixing of the $a-f_0$ and the $0^{++}$ glueball.} 
\label{fig:spec}
\end{figure}

The preliminary results we have obtained with these new parameters indicate a major influence of the lattice artifacts (see Fig.\ \ref{fig:spec}).
The most evident effect is the reduced mass of the gluino-glue. Hence a much smaller mass gap is found in the spectrum. 
A naive linear extrapolation based on the difference between the gluino-glue and the $a-\eta'$ towards the continuum limit  leads even to a negative mass gap 
at zero lattice spacing. For a precise investigation of the continuum limit we need, however, to obtain more data and investigate further improvements of the action.
\section{Conclusions}
Theoretical considerations indicate that supersymmetric Yang-Mills theory with Wilson fer\-mi\-ons on the lattice can be tuned to a supersymmetric chiral continuum limit. The simulations of this approach are challenging and advanced techniques must be used to get a reasonable signal of the observables. 
Besides the technical difficulties the theoretical considerations do not quantify the magnitude of the lattice artifacts when the continuum limit is approached. Our new simulations indicate a major influence of these artifacts on the spectrum of the theory. In particular, the mass gap between bosonic and fermionic particles at a larger lattice spacing 
is inconsistent with supersymmetry. This gap might, however, vanish in the continuum limit. Further simulations and a further refinement of the approach are needed to prove these findings.
\section*{Acknowledgments}
This project is supported by the German Science 
Foundation (DFG) under contract Mu 757/16, and by the John von Neumann 
Institute of Computing (NIC) with grants of computing time.
\begin{spacing}{0.9}

\end{spacing}

\begin{thebibliography}{99}
\bibitem{Witten:1982df}
  E.~Witten,
  Nucl.\ Phys.\  {\bf B202 } (1982)  253.

\bibitem{Veneziano:1982ah}
  G.~Veneziano, S.~Yankielowicz,
  Phys.\ Lett.\  {\bf B113 } (1982)  231.

\bibitem{Farrar:1997fn}
  G.~R.~Farrar, G.~Gabadadze, M.~Schwetz,
  Phys.\ Rev.\  {\bf D58 } (1998)  015009
  [\arxiv{hep-th/9711166}].

\bibitem{Curci:1986sm}
  G.~Curci, G.~Veneziano,
  Nucl.\ Phys.\  {\bf B292 } (1987)  555.

\bibitem{Montvay:1995ea}
  I.~Montvay,
  Nucl.\ Phys.\  {\bf B466 } (1996)  259
  [\arxiv{hep-lat/9510042}].

\bibitem{Bergner:2009vg}
  G.~Bergner,
  JHEP {\bf 1001 } (2010)  024
  [\arxiv{0909.4791} [hep-lat]].

\bibitem{Bochicchio:1985xa}
  M.~Bochicchio, L.~Maiani, G.~Martinelli, G.~C.~Rossi, M.~Testa,
  Nucl.\ Phys.\  {\bf B262 } (1985)  331.

\bibitem{Montvay:2005tj}
  I.~Montvay, E.~Scholz,
  Phys.\ Lett.\  B {\bf 623} (2005) 73
  [\arxiv{hep-lat/0506006}].

\bibitem{Koutsoumbas:1997de}
  G.~Koutsoumbas, I.~Montvay, A.~Pap, K.~Spanderen, D.~Talkenberger, J.~Westphalen,
  Nucl.\ Phys.\ Proc.\ Suppl.\  {\bf 63 } (1998)  727
  [\arxiv{hep-lat/9709091}].

\bibitem{Bergner:2011zp}
  G.~Bergner, J.~Wuilloud,
  Comput.\ Phys.\ Commun.\ (doi:10.1016/j.cpc.2011.10.007)
  [\arxiv{1104.1363} [hep-lat]].

\bibitem{Farchioni:2004ej}
  F.~Farchioni, G.~M\"unster, R.~Peetz,
  Eur.\ Phys.\ J.\  C {\bf 38} (2004) 329
  [\arxiv{hep-lat/0404004}].

\bibitem{Bali:2009hu}
  G.~S.~Bali, S.~Collins, A.~Sch\"afer,
  Comput.\ Phys.\ Commun.\  {\bf 181} (2010) 1570
  [\arxiv{0910.3970} [hep-lat]].

\bibitem{Demmouche:2010sf}
  K.~Demmouche, F.~Farchioni, A.~Ferling, I.~Montvay, G.~M\"unster, E.~E.~Scholz, J.~Wuilloud,
  Eur.\ Phys.\ J.\  C {\bf 69} (2010) 147
  [\arxiv{1003.2073} [hep-lat]].
\end{thebibliography}
\end{document}